# Inverse Design of Promising Alloys for Electrocatalytic $CO_2$ Reduction via Generative Graph Neural Networks Combined with Bird Swarm Algorithm

Zhilong Song[1#], Linfeng Fan[2,1 #], Shuaihua Lu[1], Qionghua Zhou[1,2,*], Chongyi Ling[1*], and Jinlan Wang[1,2,*]

[1]Key Laboratory of Quantum Materials and Devices of Ministry of Education, School of Physics, Southeast University, Nanjing 21189, China

[2] Suzhou Laboratory, Suzhou, China

[#]These authors contributed equally: Zhilong Song and Linfeng Fan

Directly generating material structures with optimal properties is a long-standing goal in material design. One of the fundamental challenges lies in how to overcome the limitation of traditional generative models to efficiently explore the global chemical space rather than a small localized space. Herein, we develop a framework named MAGECS to address this dilemma, by integrating the bird swarm algorithm and supervised graph neural network to effectively navigate the generative model in the immense chemical space towards materials with target properties. As a demonstration, MAGECS is applied to design compelling alloy electrocatalysts for $CO_2$ reduction reaction ($CO_2$RR) and works extremely well. Specifically, the chemical space of $CO_2$RR is effectively explored, where over 250,000 promising structures with high activity have been generated and notably, the proportion of desired structures is 2.5-fold increased. Moreover, five predicted alloys, *i.e.*, CuAl, AlPd, $Sn_2Pd_5$, $Sn_9Pd_7$, and $CuAlSe_2$ are successfully synthesized and characterized experimentally, two of which exhibit about 90% Faraday efficiency of $CO_2$RR, and CuAl achieved 76% efficiency for $C_2$ products. This pioneering application of inverse design in $CO_2$RR catalysis showcases the potential of MAGECS to dramatically accelerate the development of functional materials, paving the way for fully automated, artificial intelligence-driven material design.

## 1. Introduction

Machine learning (ML), owing to its ability to analyze vast datasets and identify complex correlations, has revolutionized the material science landscape[1–4]. Currently, a widely-used ML approach, known as forward design, involves utilizing ML algorithms to establish structure-property relationships and predicts properties of unknown materials through element substitution within existing materials[5–9]. However, it is extremely challenging to identify novel and useful candidates in search spaces that are overwhelming in size, *e.g.*, the chemical space for drug-like molecules is estimated to contain >$10^{33}$ structures, not to mention more complex inorganic materials[10]. Furthermore, a fundamental drawback of forward design is that it cannot generate materials beyond structural prototypes of existing materials. In recent years, generative models, particularly those based on deep learning architectures, have emerged as powerful tools to overcome the drawbacks of forward design for discovering materials (Figure S1)[11,12]. These models leverage large datasets of materials properties and structures to learn complex patterns and correlations. By capturing the intricate interplay between various material attributes, generative models can generate new materials that meet specific criteria. This strategy is named inverse design[11]. This transformative capability not only accelerates the discovery process but also expands the scope of materials exploration beyond what was previously conceivable[13].

Two popular generative models, the variational autoencoder (VAE) and the generative adversarial network (GAN) have been successfully applied to inversely design stable V-O[14], Bi-Se[15], Mg−Mn−O[16] material systems, zeolites with desired methane heat of adsorption[17], stable cubic semiconductors[18,19] and MOFs for carbon dioxide separation[20]. However, these generative models are often limited to the generation of structures with a given symmetry or composition. Most recently, the crystal diffusion variational autoencoder (CDVAE) based on graph neural networks (GNN) has been developed to generate diverse crystal structures with accepted quality.[21] Nevertheless, generative models inherently learn patterns from large and diverse training datasets to achieve the generation of high-quality structures.

Consequently, they tend to generate material structures that closely resemble the training dataset, where material structures with superior properties are usually rare. This suggests we need additional algorithms to guide the generative model to escape away from the restriction of the training data and simultaneously maintain the generation quality. Meanwhile, the immensity of the material property space (*e.g.*, $10^{33}$) presents significant challenges for global and efficient exploration. On the other hand, the current inverse design methods used for designing materials with targeted properties lack universality, necessitating the training of different generative models for different properties[14–16,18,19,21]. This methodology fails to leverage a plethora of accurate property prediction models already available, which use various material descriptors, both compositional and structural[6,7,9,22–25]. As a result, the capability of inverse design is restricted to a narrow set of properties, primarily formation energy.

In this work, we address the aforementioned challenges by developing a general inverse design framework named MAGECS (Material Generation with Efficient global Chemical space Search), which integrates bird swarm algorithm (BSA)[26,27], crystal diffusion variational autoencoder (CDVAE) and supervised graph neural network (GNN). The introduction of BSA can efficiently steer the generator towards generating structures with target property via optimizing latent space vectors, which serve as the input for generative models to construct structures, in the property space. This transforms the generation of structures from the traditional random generation to purposeful and efficient exploration of property space based on targeted properties. Using MAGECS, we realize the first-ever inverse design of novel alloy electrocatalysts for $CO_2RR$—a pivotal step in mitigating greenhouse gas emissions and promoting the carbon cycle[28–32]. To effectively evaluate the $CO_2RR$ activity of alloys, we utilize the optimal adsorption energy of CO ($\Delta E_{CO}$), which is usually the key intermediate in electrocatalytic $CO_2RR$[8,33]. Out of the 250,000 alloy surfaces we generated, the proportion of structures with high $CO_2RR$ activity is 2.5 times higher than structures generated by conventional CDVAE. Next, among these highly active alloy surfaces, we further consider the competitive hydrogen evolution reaction (HER) and thermodynamic stability, and screen the top 110 potential surfaces for further verification with first-principles calculations. To the end, we successfully synthesized five innovative alloy catalysts, CuAl, AlPd, $Sn_2Pd_5$, $Sn_9Pd_7$, and $CuAlSe_2$, and two of which exhibit around 90% $CO_2RR$ Faraday efficiency.

## 2. Results and discussion

### 2.1. Inverse design framework of MAGECS

Our inverse design framework MAGECS comprises two primary domains of operation: the generation of new surfaces in structure space and the global optimization of generated surfaces in property space (Figure 1). First, we employ the CDVAE pretrained generative model to create new surfaces with both high quality and diversity. The CDVAE model is trained on a database containing various catalyst surfaces created by Tran *et al.* (GASpy) (Table S1) and generates new structures (*i.e.*, surfaces in this work) from latent vectors (steps I-II in Figure 1). Specifically, these latent vectors are fed into three fully connected neural networks which respectively output the atomic number, crystal structure, and composition (Figure S2). A surface can be built using this information and an additional set of randomly generated atom coordinates. During surface generation, the Langevin dynamics is employed to ensure the reasonability of the atom coordinates and compositions.

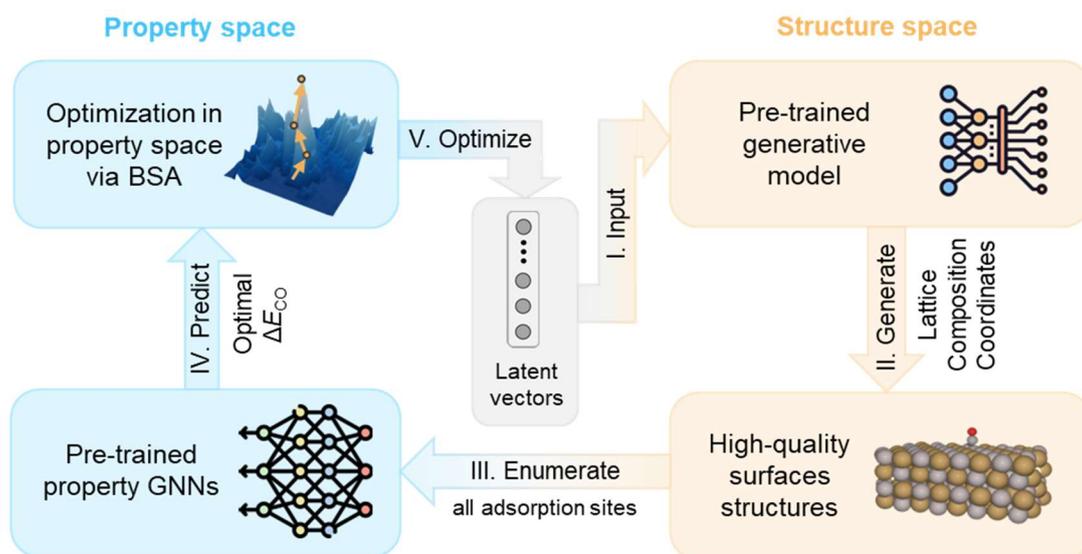

**Figure 1.** Schematic diagram of inverse design materials with desired properties using our framework (surfaces for $CO_2RR$ in this work). It contains three main parts: surface structures generation from latent vectors via generative model (step I-II), CO adsorption energy prediction

via supervised GNN model (step III-IV) and optimization of $CO_2RR$ properties via BSA (step V).

Second, to realize the optimization of generated surfaces in property space, we need to rapidly assess the $CO_2$ activity of these generated surfaces. Here the adsorption energy of the key intermediate CO ($\Delta E_{CO}$) is selected as the learning target. To this end, we enumerate all possible adsorption sites on the surface and add CO on these sites (step III in Figure 1). We then employ a pre-trained supervised graph neural network (DimeNet++), recognized for its invariance to crystal structures and excellent accuracy in predicting material properties[34,35], to predict the adsorption energies of all sites. Next, the minimal predicted adsorption energies of all sites are used as fitness for evaluating latent vectors (step IV in Figure 1). It is noteworthy that our framework is compatible with any form of property prediction model, regardless of the type of material descriptors they utilize, whether based on composition or structure. This enables our framework to optimize any material properties without altering the generative model.

Third, to steer the generator toward generating active surfaces for $CO_2RR$ (*i.e.*, to globally optimize the latent vectors in the property space), we integrate the BSA algorithm which was inspired by the swarm intelligence observed in bird swarms. As depicted in Figure S2, birds in nature exhibit three main social behaviors: foraging, vigilance, and flight. By modeling these interactions, BSA not only exhibits superior optimization efficiency but also has excellent capability of escaping from local optimum. In this work, BSA first generates a batch of birds (*i.e.*, latent vectors in Figure S3), which can be used to generate an equal number of surface structures. The activity of these surfaces is evaluated by DimeNet++ and then fed back to BSA to generate new latent vectors (step V in Figure 1). This process will be iterated until a number of predetermined generations are reached.

To sum up, the BSA and supervised models are employed to guide (arrows in Figure 2a) the optimization of the CDVAE-generated structures in the property space, enabling rapid discovery of new structures beyond the training database (blue points in Figure 2a) with excellent properties (peaks in Figure 2a). As a proof to this, Figure 2b reveals that our framework indeed explores a much larger chemical space with high

CO$_2$RR activity than that covered by the training database. Notably, this training database, deriving from active learning methods and automated DFT calculations[8], already covers a vast chemical space. This further proves the superiority of our inverse design framework in globally and efficiently exploring the chemical space.

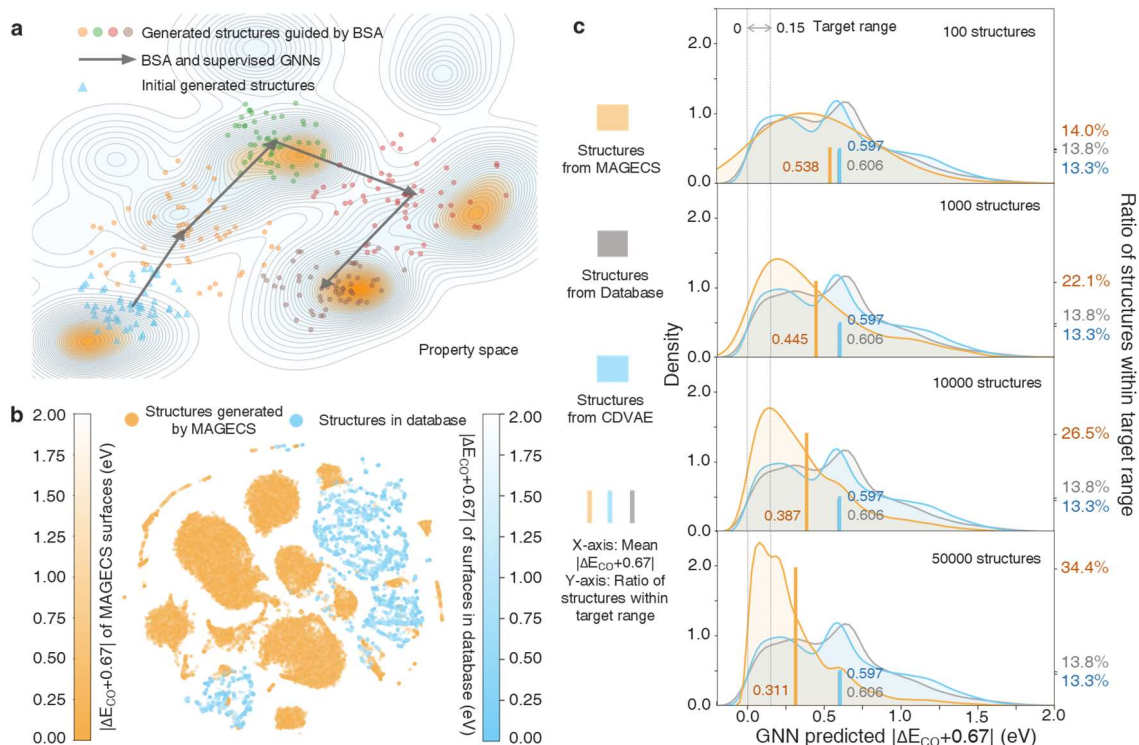

**Figure 2. (a)** Schematic diagram of the superiority of our inverse design framework. The structures generated by CDVAE (dots) are optimized by BSA and supervised models (arrows), which enables the global and efficient exploration of peaks (materials with good properties) in the property space. **(b)** T-SNE visualization of the structures in dataset (blue points) and MAGECS generated structures (orange points) using our framework. The darker the color of the point, the better the CO$_2$RR activity. **(c)** Comparison of the distribution of $|\Delta E_{CO} + 0.67|$, the average value of $|\Delta E_{CO} + 0.67|$ (X-axis of columns) and the proportion of structures satisfying $|\Delta E_{CO} + 0.67| \leq 0.15$ eV (Y-axis of columns) among MAGECS, CDVAE generated structures and training set structures. From top to bottom, the four figures illustrate the comparison where MAGECS generated 100, 1,000, 10,000, and 50,000 structures, respectively.

## 2.2. Results of inverse design framework and evaluation of generated surfaces

To realize the global optimization of generated surfaces for CO$_2$RR in the property space, the optimization target is first required to be set. According to previous studies, the optimal $\Delta E_{CO}$ for CO$_2$RR determined by microkinetic modeling is -0.67 eV[8,33]. Considering the mean absolute error (MAE) of the DimeNet++ model (0.143 eV on testing data), we set the criterion with $|\Delta E_{CO} + 0.67| \leq 0.15$ eV to be favorable and used it as the optimization target for BSA. MAGECS was then executed three times, each comprising 500 BSA steps (100 structures each step). In all three runs, the BSA can lead the generative model to go beyond the training data (Figure S4), demonstrating the effectiveness of MAGECS. Note that there is no further improvement in the number of generated surfaces meeting $|\Delta E_{CO} + 0.67| \leq 0.15$ eV after 200-300 BSA steps. In order to validate this finding and generate more promising surface structures with high $CO_2$RR activity, one additional run with 1000 BSA steps was conducted. Hence, a total of 250,000 alloy surfaces were generated.

To reveal the advantage of MAGECS, we employed the conventional CDVAE model to produce 50000 new surfaces and compare the distribution of predicted $|\Delta E_{CO} + 0.67|$ across surfaces generated by MAGECS, CDVAE, and those within the training set. As shown in Figure 2c, the CDVAE successfully reproduced the distribution of $|\Delta E_{CO} + 0.67|$ of training surfaces, with both the average $|\Delta E_{CO} + 0.67|$ and proportion of surfaces satisfying $|\Delta E_{CO} + 0.67| \leq 0.15$ eV (highly active surfaces) closely aligning. In contrast, among the 100, 1000, 10000 and 50000 structures iteratively generated by MAGECS, the proportion of highly active surfaces was rapidly improved, ultimately being 2.5 times higher than those generated by CDVAE and from training data. The above merits showcase the efficacy of our BSA to steer the generative model to mass-generate structures with properties beyond training data.

Building upon this, we conducted a thorough analysis of BSA optimization process using the average performance of four runs (Figure S4-S6 detail the results of each run) and the mean value of every ten BSA steps considering the deviation during optimization. The efficiency of the MAGECS framework is highlighted in Figure 3a, where we observed the proportion of surfaces with $|\Delta E_{CO} + 0.67| \leq 0.15$ eV out of 100 generated surfaces in each step increases rapidly as BSA runs, eventually maintaining around 38% after 200 steps, which yields a considerable number of desired surfaces. Meanwhile, 78.9% of the generated surfaces exceed the training and validation sets (subfigure of Figure 3a), proving the powerful capability of MAGECS to create new materials with enhanced properties. Moreover, as shown in Figure 3b, a distinctive feature of the BSA optimization process is the approximately 55% similarity between the surfaces generated in successive BSA steps. This continuous generation of novel and superior surfaces, even after identifying those with the lowest $|\Delta E_{CO} + 0.67|$, demonstrates MAGECS's capability to transcend local minima and undertake a global exploration of the chemical space. More importantly, among the generated surfaces, we found a number of them have been experimentally verified to exhibit excellent $CO_2RR$ performance in previous studies (Figure 3b)[3,36–43]. As shown in Figure S7, these rediscovered surfaces indeed have low $|\Delta E_{CO} + 0.67|$, demonstrating the reliability of MAGECS in generating highly active surface structures for $CO_2RR$.

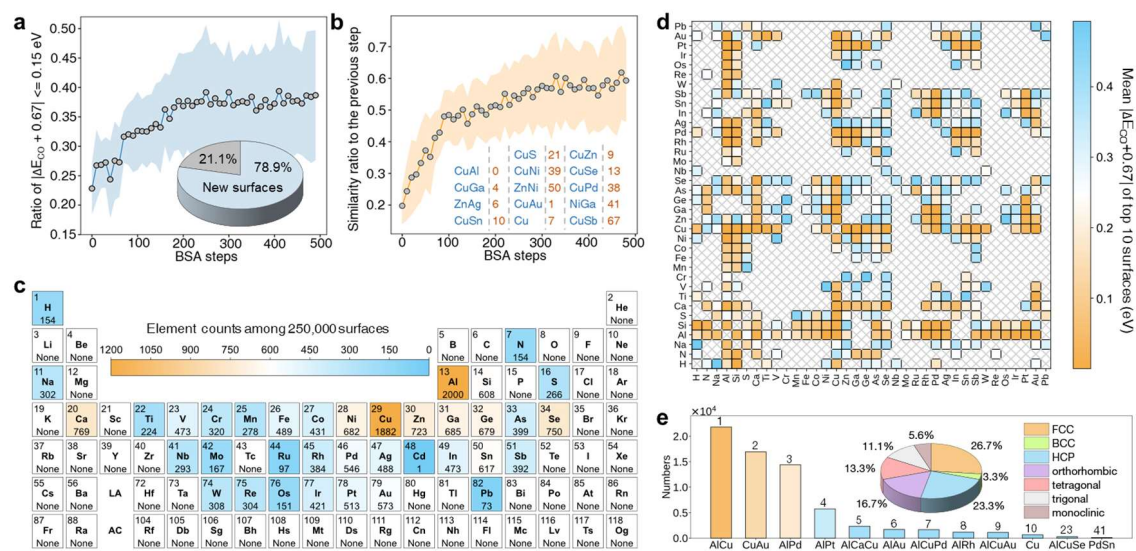

**Figure 3. (a)** Average proportion of surfaces with $|\Delta E_{CO}+0.67| \leq 0.15$ eV measured at every ten steps during four runs of BSA optimization. The error band indicates standard deviation. Inset displays the proportion of new surfaces among 250,000 generated surfaces. **(b)** Mean surface similarity ratio in every ten steps during BSA optimization of four runs. Inset shows the number of steps rediscovering formerly reported surfaces with high $CO_2RR$ performance. **(c)** Preferences of elements across 250,000 generated surfaces, with increasing orange intensity indicating the greater quantity of element. **(d)** Predicted activity distribution for generated bimetallic alloy surfaces. The more orange the color, the lower the average of the lowest ten $|\Delta E_{CO}+0.67|$ of surfaces. **(e)** Top composition of 250,000 generated surfaces with $|\Delta E_{CO}+0.67| \leq 0.15$eV. The number on the bar represents the rank, the subfigure shows the crystal system distribution of the bulk structure of these surfaces.

Although the generated structures exhibit commendable $CO_2RR$ activity, evaluating the rationality of them is imperative. Formation energy, as a qualitative description of thermodynamic stability, has been demonstrated to be a suitable structure evaluation metric[44]. In order to avoid time-consuming DFT calculations and accelerate the prediction capability, we adopted a high-precision graph neural network model (MEGNet) to predict the formation energy. The proof of the rationality of our structure evaluation metric are discussed in Supplementary Methods 1 and Figure S8-13. As shown in Figure S8c, the formation energies of generated surfaces predicted by MEGNet has a mean value and distribution closely matching those of the training and validation surfaces. This indicates the capability of generating high-quality structures of the generation model in our inverse design framework.

After validating the rationality of 250,000 generated surfaces, an in-depth analyze of the elemental, compositional, and structural distributions across these surfaces helps identify key factors in $CO_2RR$. We first visualized the frequency of occurrence for each element across the 250,000 alloy surfaces (Figure 3c), where Cu and Al are emerged as the most prevalent elements. To further elucidate combinations of elements favorable for $CO_2RR$, we computed the average of the smallest ten $|\Delta E_{CO} + 0.67|$ values for binary alloys (Figure 3d). Overall, binary alloys comprising Cu and Al in combination with other elements exhibit the most favorable $CO_2RR$ activity, alongside numerous high-activity binary alloy surfaces yet to be explored, such as AlPt, AlPd, and SnPd. Beyond binary alloys, 250,000 surfaces include pure metals, ternary and quaternary alloys, with 1,549 out of a total of 4,573 compositions satisfying $|\Delta E_{CO} + 0.67| \leq 0.15$ eV. Figure 3e showcases the foremost ten compositions, with AlCu, CuAu, AlPd, and AlPt topping the list, meanwhile highlighting promising yet unexplored compositions like CuAlSe (23rd) and SnPd (41st). These statistical analyses on elements and compositions offer significant guidance for designing $CO_2RR$ alloys and align with experimental evidence. Specifically, it is well-known that Cu is the most used element for $CO_2RR$ and various Cu-based alloys like CuAl, CuAu, CuPd, CuGa have demonstrated the exceptional ability in reducing $CO_2$ to diverse products[3,37,45–49]. CuAl alloy, in particular, has been experimentally demonstrated to have the state-of-the-art Faraday efficiency in producing ethylene as well[3]. Moreover, among the surfaces meeting the $|\Delta E_{CO} + 0.67| \leq 0.15$ eV, those in close-packed arrangement with face-centered cubic (FCC) or hexagonal close-packed (HCP) phase constitutes the majority, accounting for 26.7% or 23.3%, respectively. Meanwhile, due to the presence of main group elements, some non-close-packed structures have also been generated, including orthorhombic, tetragonal phases, etc. These results support the reliability of our framework in generating stable structures once again.

## 2.3. Identification of suitable surfaces for CO₂RR via DFT calculations

Above our framework has successfully generated a large number of rational and potential surfaces with excellent CO$_2$RR performance. Specifically, 89,875 surfaces satisfy $|\Delta E_{CO} + 0.67| \leq 0.15$ eV (the first selection criterion). Next, considering the limitation of computational resources, we try to select the best few surfaces among these surfaces for DFT verification. During the CO$_2$RR, the HER needs to be suppressed because it competes for active sites on surfaces with CO$_2$RR. Given that the lowest limiting potential of HER occurs when $\Delta G_H = 0$ ($\Delta E_H = \Delta G_H - 0.27 = -0.27$ eV)[50], which is most conducive to the HER, the $|\Delta E_H + 0.27|$ should be as high as possible. On the other hand, $\Delta E_H$ should not be too negative, otherwise the *H intermediate will occupy too many active sites and cause catalyst poisoning. Thus, the second selection criterion is set as $\Delta E_H + 0.27 \geq 0.6$ eV. In addition, to make the generated surface as reasonable as possible, we set the third criterion to ensure that the MEGNet predicted formation energy of the generated surfaces $\leq -0.1$ eV/atom.

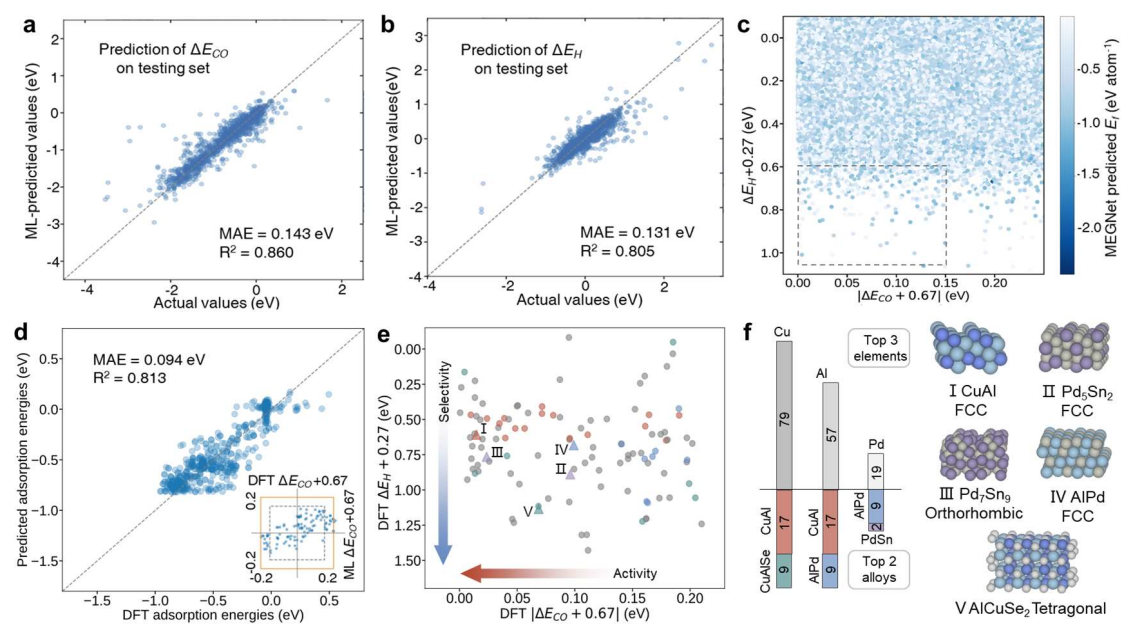

**Figure 4.** DimeNet++ predicted **(a)** CO and **(b)** H adsorption energies vs DFT calculated adsorption energies on independent testing data. **(c)** $|\Delta E_{CO}+0.67|$ vs $|\Delta E_H+0.27|$ of 250,000 surfaces, the color represents the value of MEGNet predicted formation energy. 110 surfaces selected for DFT validation are in the rectangle area. **(d)** DimeNet++ predicted $\Delta E_{CO}$ vs DFT calculated $\Delta E_{CO}$ on 110 surfaces selected for DFT validation. The subfigure shows range of

ML-predicted and DFT-calculated minimum $\Delta E_{CO}$, the orange and grey rectangle border ranges from -0.2 to 0.2 eV and -0.15 to 0.15 eV, respectively. **(e)** DFT calculated $|\Delta E_{CO}+0.67|$ vs $|\Delta E_H+0.27|$ of 110 surfaces. CuAl, AlPd, SnPd, CuAlSe and other surfaces are marked with red, blue, purple, green and grey points, respectively. Five surfaces selected for experimental validation are in the rectangle area and marked with triangle. **(f)** Top 3 metal elements among 110 surfaces and top 2 alloys of Cu, Al, Pd-based surfaces. Structure, crystal system of bulk and miller index of the five synthesized surfaces.

The above three selection criteria for stable surfaces with high stability, activity and selectivity for CO$_2$RR were achieved by accurate GNNs. The adsorption energies in the first two criteria were predicted by two pre-trained DimeNet++ models. As depicted in Figure 4a, b, the DimeNet++ models accurately predict CO and *H adsorption energies on the independent test set with MAEs of 0.143 eV and 0.131 eV, and $R^2$ of 0.860 and 0.805, respectively. Considering the complexity of the training data (13,000 various structures), our models' accuracy is satisfactory and outperforms both the random forest model and GNN model in the literature[8,51] that used the same data set. The hyperparameters and training details of our DimeNet++ models are shown in Table S2 and Figure S14-15, respectively. As for the third selection criterion, the pre-trained MEGNet model for the formation energy achieves an MAE of 0.017 eV/atom on the independent test data after being trained with data from the OQMD dataset[44]. With the help of three GNNs, 110 best surfaces were selected based on the three selection criteria and are shown as dots in the rectangle area in Figure 4c. Similar to the initial 250,000 surfaces, the most commonly used elements of 110 surfaces are still Cu and Al (Figure S16), and the most frequent element combinations are AlCu, AlCuSe and AlPd (Figure S17). Notably, after considering selectivity, the unexplored AlCuSe and the non-copper-based alloys SnPd, AlAu, and SbPt have emerged among the top 15 leading compositions.

Next, we carried out DFT verification on the selected 110 surfaces. First, all possible adsorption sites were enumerated on each surface, then DFT calculations were performed to determine the adsorption energies after adding CO and *H species. The most negative adsorption energies across different sites on each surface were taken as the final adsorption energies. As shown in Figure 4d, our DFT-calculated $\Delta E_{CO}$ (a total of 1385 calculations on all sites) show good agreement with those predicted by the

DimeNet++ model, with an MAE of 0.094eV and $R^2$ of 0.813, even though differences may exist in calculation parameters between our calculations and previous literature[8]. More importantly, the subfigure of Figure 4d evidently proved the effectiveness of our framework: 97% of generated surfaces have DFT-calculated minimum $|\Delta E_{CO} + 0.67| \leq$ 0.2 eV (the maximum value is 0.210 eV), which is five times higher than that of training data (19.6%). As a result, the distribution of $|\Delta E_{CO} + 0.67|$ of generated surfaces is much closer to 0 eV than that of surfaces in the database (Figure S18).

**2.4. Identification of suitable surfaces for CO$_2$RR via experiments**

Finally, we select promising alloys for experimental validation. Considering synthesizability, our selection was narrowed to alloys composed of three or fewer elements. The compositions frequently observed across the screened 110 surfaces are promising for superior CO$_2$RR activity and selectivity. Thus, we first identified the most prevalent metal elements among 110 surfaces: Cu, Al, and Pd. Subsequently, we focused on surfaces primarily composed of Cu, Al, and Pd, selecting the top two compositions for each (Figures 4e and f). As a result, CuAl, AlPd, Sn$_2$Pd$_5$, Sn$_9$Pd$_7$, and CuAlSe$_2$ were chosen, with their surface structures and bulk crystal systems illustrated in Figure 4f. Comprehensive characterization and performance tests were conducted on CuAl and Sn$_2$Pd$_5$.

The Scanning electron microscopy (SEM) images (Figure S19) and transmission electron microscopy (TEM) images (Figure 5a) show that CuAl alloys exist as amorphous nanoblocks. The high-resolution TEM (HRTEM) image of the CuAl (inset in Figure 5a) displays clear lattice fringes with a lattice distance of 0.2 nm corresponding to the (411) plane of the prepared catalyst. The energy dispersive spectroscopy (EDS) elemental mapping analysis indicates a homogeneous distribution of the two elements in the alloys without significant phase separation (Figure 5a). The powder X-ray diffraction (XRD) pattern (Figure 5b) confirmed the successful synthesis of the alloy. Surface elemental composition of CuAl alloys was investigated through X-ray photoelectron spectroscopy (XPS) measurements, revealing the presence of Cu, Al, and O elements on the surface (Figure S20a). High-resolution Cu LMM and Cu 2p spectra indicate the presence of metallic Cu and copper oxide (Figure S20b and c).

Figure S20d illustrates the presence of metallic and oxidized aluminum states on the surface of the alloy. The presence of the atmosphere leads to partial oxidation of the surface of the synthesized nano-alloys. We evaluated the performance of CuAl in electrochemical $CO_2RR$ using the timed-current method in a flow-type cell equipped with a three-electrode system in a 1 M KOH solution. Gaseous and liquid products were analyzed using online gas chromatography and nuclear magnetic resonance (NMR) spectroscopy. The presented data demonstrate the excellent performance of CuAl catalysts in $CO_2RR$ reactions (Figure 5c and 5g). And at -0.7 V vs. RHE, the catalyst exhibited the highest $CO_2RR$ performance, achieving an overall Faraday efficiency (FE) of 87.73% (Figure 5c and S21). Furthermore, the catalyst retained a relatively high overall FE after 24 hours of continuous electrolysis at -0.7 V vs. RHE (Figure 5h), presented an outstanding electrochemical stability.

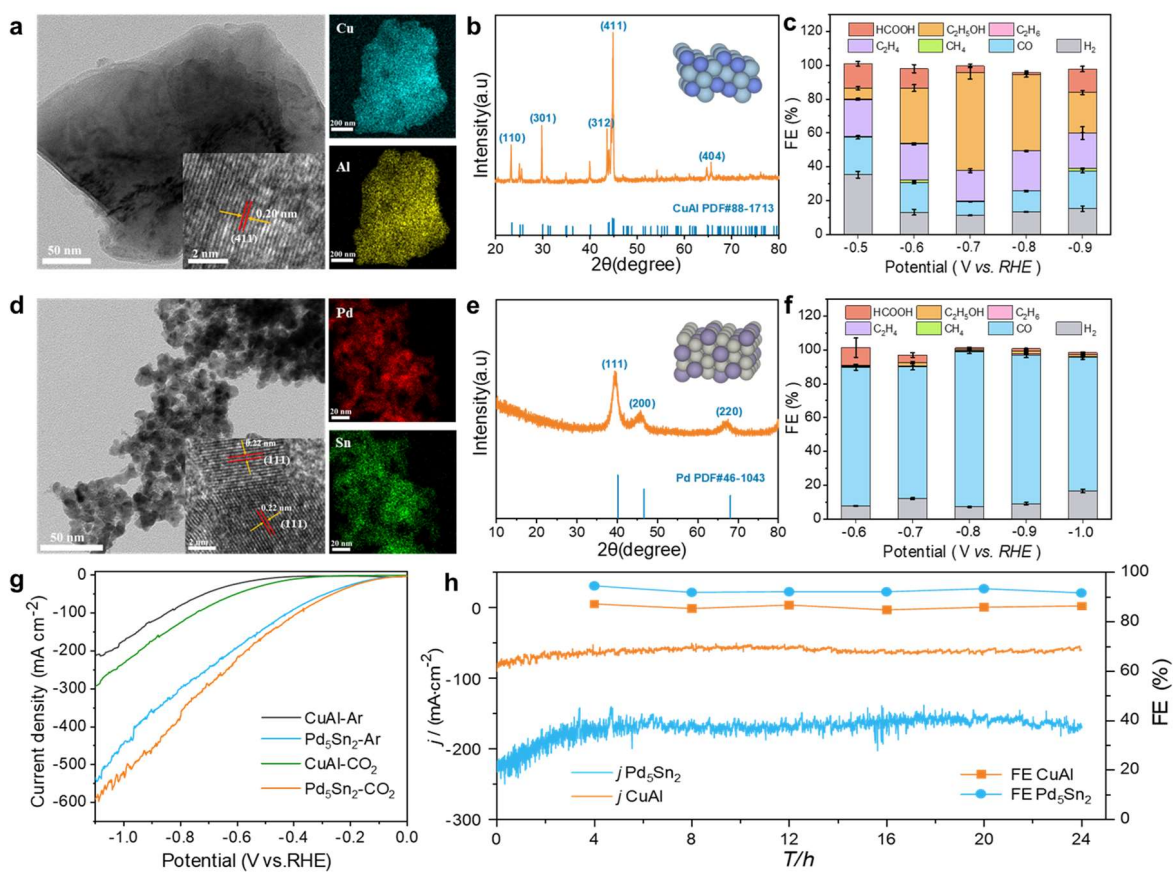

**Figure 5**. **(a)** TEM images and EDS elemental mapping images of Cu, Al for CuAl alloys (inset: HRTEM image). **(b)** XRD patterns of CuAl alloys (subfigure visualizes the generated CuAl surface). **(c)** The FEs towards $CO_2RR$ products under a range of applied potentials under 1 M KOH of CuAl alloys. **(d)** TEM images and EDS elemental mapping images of Pd, Sn for $Pd_5Sn_2$

alloys (inset: HRTEM image). **(e)** XRD patterns of $Pd_5Sn_2$ alloys (subfigure visualizes the generated $Pd_5Sn_2$ surface). **(f)** FEs towards $CO_2RR$ products under a range of applied potentials under 1 M KOH of $Pd_5Sn_2$ alloys. (g) LSV curves of CuAl alloys and $Pd_5Sn_2$ alloys. **(h)** Electrochemical stability test of the CuAl catalyst at -0.7 V vs. RHE and $Pd_5Sn_2$ catalyst at -0.8 V vs. RHE (FE is the total of the $CO_2RR$ to carbon products).

In contrast, the $Pd_5Sn_2$ materials synthesized through wet chemistry comprise agglomerated nanoparticles (Figure S22). TEM images also indicate that the alloy particles exhibit a uniform size of around 10 nm (Figure 5d). HRTEM images showed the lattice spacings of 0.22 nm (inset in Figure 5d), corresponding to the (111) planes of Pd. Additionally, EDS elemental mapping analysis demonstrates the uniform distribution of Pd and Sn across the chosen area (Figure 5d). The XRD pattern illustrates that $Pd_5Sn_2$ displays a pattern similar to that of Pd (JCPDS No. 46-1034), lacking peaks attributed to Sn-based compounds. However, a shift to lower angles, in comparison to the pattern of Pd, suggests the uninformed doping of Sn into Pd (Figure 5e). The survey XPS spectrum in (Figure S23) validates the presence of Pd and Sn elements on the nanoparticle surface. The assessment of $Pd_5Sn_2$ in $CO_2RR$ was conducted under identical conditions. The catalyst achieved a faradaic efficiency exceeding 80% for the conversion of $CO_2$ to carbon products within the voltage range of -0.6 to -1.0 V vs RHE. Moreover, the faradaic efficiency (FE) for CO production remained steady at approximately 80% across the potential window, consistently generating CO with an average FE as high as 91.86% at −0.8 V vs RHE (Figure 5f and S24). And $Pd_5Sn2$ maintained good stability in terms of FE of ~90% for $CO_2RR$ to carbon products at -0.8 V vs. RHE, remaining stable for 24 hours. In addition, $Pd_7Sn_9$, PdAl, and $CuAlSe_2$ were synthesized and all exhibited good $CO_2RR$ properties. The FEs for $CO_2RR$ to carbon products at specific voltages were all above 70% (Figure S25-35).

## Conclusion

In summary, we have developed a general property-to-structure inverse design framework, MAGECS, which enables comprehensive exploration of vast chemical spaces and consistent generation of high-quality material structures with target properties. This merit has been realized by innovatively integrating the bird swarm algorithm with state-of-the-art GNNs to effectively navigate the generative model toward materials with superior properties. The efficiency of MAGECS has been robustly demonstrated in the application of designing alloy surfaces for electrocatalytic $CO_2RR$. A total of 250,000 rational and promising alloy surfaces were generated and 110 surfaces were subjected to first-principles calculations, due to their high predictive activity, selectivity, and stability. Significantly, the ratio of surfaces exhibiting high activity for $CO_2RR$ surpasses the training data benchmarks by a remarkable 2.5 times. On this basis, we synthesized and characterized five novel alloys—CuAl, AlPd, $Sn_2Pd_5$, $Sn_9Pd_7$, and $CuAlSe_2$. Among these, two alloys demonstrated approximately 90% Faraday efficiency in $CO_2RR$, with CuAl notably achieving 76% efficiency for $C_2$ products.

While our developed MAGECS demonstrates considerable capability in the inverse design of $CO_2RR$ electrocatalysts, there is still potential for further enhancements. Specifically, the accuracy of supervised GNN models presents room for improvement, as it currently limits the efficiency of our framework. Moreover, numerous functional materials, such as photocatalysts, require the simultaneous satisfaction of multiple target properties including band gap, band edge, and stability. This necessitates the development of efficient multi-objective optimization strategies integrated with inverse design. Finally, considering the gap between thermodynamic stability and experimental synthesizability, integrating a universal model to predict synthesizability into our framework will undoubtedly catapult MAGECS to a greater height.

## 3. Methods

### 3.1. Crystal diffusion variational autoencoder

CDVAE consists of two GNNs (GemNet and DimeNet++ are used in this work) and three fully connected neural networks (NNs). While training, the DimeNet++ was trained to encode the original material structures into latent vectors and the GemNet was trained to denoise the noised material structures. The main hyperparameters for training CDVAE are shown in Table S4-6. All main loss functions on training set and validation set converge well (Figure S36), demonstrating the training of CDVAE was adequate.

**3.2. Supervised GNN**

Six popular supervised GNNs (CGCNN[52], SchNet[53], DimeNet++[34,35], PaiNN[54], GemNet[55] and GemNet-OC[51]) were tested for surface property prediction. As shown in Figure S37, the CGCNN and SchNet, which did not consider the information of interatomic angles, showed a clear performance gap with other GNNs on testing data. The DimeNet++ had comparable performance to PaiNN and outperformed the GemNet and GemNet-OC on testing data. Moreover, the DimeNet++ exhibited faster training speed than PaiNN (1.2x), GemNet (2x) and GemNet-OC (2.5x). Thus, the DimeNet++ was utilized to predict CO and H adsorption energies in this work. The main hyperparameters of training these GNNs are listed in Table S2, 7-11.

**3.3. Bird swarm algorithm**

The BSA algorithm was inspired by the swarm intelligence observed in bird swarms. Birds exhibit three main behaviors: foraging, vigilance, and flight. These social interactions help birds find food and avoid predators, increasing their chances of survival. BSA models these behaviors with five simplified rules (see Supplementary methods 2 and Figure S2), endowing it with excellent optimization efficiency and the ability to escape local optima. Thus, the BSA is used for global exploration of the chemical space in our framework. Table S4 shows the hyperparameters of performing BSA. We wrote the BSA with python language and the code is available in https://github.com/szl666/inverse_design.

**3.4. Automated DFT calculations**

All the DFT energies in this work were calculated by the VASP package[56]. The slabs with all kinds of adsorbates were automatically built using the Pymatgen[57] package. The structure optimization was performed using a plane wave-based group with a 350 eV cutoff energy and the RPBE exchange-correlation function. The energy and force convergence criterion are set to be $5 \times 10^{-4}$ eV and 0.02 eV/Å, respectively.

The Gibbs free energy difference ΔG of the adsorbates before and after adsorption on the surface was calculated to consider temperature's influence. The standard hydrogen electrode approximation was used to bypass electron energy calculation. The free energy was calculated by Eq. (1):

$$G = E + ZPE – TS \qquad (1)$$

where *E* is energy, *ZPE* and *S* are the zero-vibration energy and entropy, respectively, which can be obtained by VASP vibrational calculation. *T* is the temperature setting to 300 K.

### 3.5. Material synthesis

All chemicals are of analytical grade and used without further purification. Copper powder, aluminum powder and palladium powder were purchased from Hebei Jiuyue New Material Technology Co. Palladium diacetylacetonate and selenium powder were bought from Aladdin. $SnCl_2$, $NaBH_4$, Polyvinyl pyrrolidone (PVP) and ethylene glycol were bought from Mackli. Ethanol, potassium format, $N_2H_4·H_2O$ and KOH were purchased from Sinopharm. The ion exchange membrane was purchased from Dupont. Carbon paper (CP) was purchased from Avcarb. The detail of the synthesis of CuAl, $CuAlSe_2$, PdAl, $Pd_{15}Sn_6$ and $Pd_{15}Sn_6$ are discussed in Supplemental Methods 3. The EDS results of above-mentioned materials are shown in Table S12.

### 3.6. Material characterization

The phase and crystallinity of samples were characterized by X-ray diffraction (Miniflex6000, Rigaku) at 40 kV and 15 mA using Cu-Kα radiation (λ =1.54178 Å) at room temperature and scan speed was 15°/min. Morphology of the catalysts was characterized by High resolution field emission scanning electron microscope (FEI Inspect F50) and Thermo Scientific Talos F200X transmission electron microscope (STEM, Talos F200X). X-ray photoelectron spectroscopy (XPS) was performed on Escalab 250Xi. NMR spectra were recorded on a AVANCE III HD 600MHz. In which 500 μL electrolyte was added with 100 μL D2O and dimethyl sulfoxide (DMSO) was added as the internal standard.

### 3.7. Electrochemical characterization

The flow cell assembly used consists of a gas flow chamber, a cation chamber and an anion chamber. Each chamber has an inlet and outlet for electrolyte or gas. A commercial platinum sheet (0.5*0.5 cm$^2$) is used as anode and an Ag|AgCl is acted as the reference. 1 M KOH are used as the cathode and anode electrolyte, respectively. The fabrication process for the working electrode involved adding 100 μL of catalyst ink dropwise to the carbon paper (0.5*0.5 cm$^2$) electrode to achieve a loading of approximately 1 mg cm$^{-2}$. The electrode was then dried under an infrared lamp. The cathode chamber and anode chamber were separated by a piece of ion exchange membrane. Electrolytes were cycled at 20 mL min$^{-1}$ and the CO$_2$ gas was supplied at rate of 20 sscm.

All potentials in our experiments are converted to reversible hydrogen electrode (RHE) reference scale by using the Nernst function as below,

E (*vs.* RHE) = E (*vs.* Ag|AgCl) + 0.197 V + 0.059 × pH

The detail about the products analysis is discussed in Supplementary Methods 4 and Figure S34-35.

**Acknowledgements**

This work was supported by the National Key Research and Development Program of China (grant 2022YFA1503103, 2021YFA1200700), the Natural Science Foundation of China (grant 22033002, 9226111, 22373013, T2321002), and the Basic Research Program of Jiangsu Province (BK20232012, BK20222007). We thank the National Supercomputing Center of Tianjin and the Big Data Computing Center of Southeast University for providing the facility support on the calculations.